\documentstyle[12pt]{article}
\begin{document}

\title{Rotating vacuum wormhole}
\author{V.M.Khatsymovsky \\
 {\em Budker Institute of Nuclear Physics} \\ {\em Novosibirsk,
 630090,
 Russia} \\ {\em E-mail address: khatsym@inp.nsk.su}}
\date{}
\maketitle
\begin{abstract}
We investigate whether self-maintained vacuum traversible
wormhole can exist described by stationary but nonstatic
metric.  We consider metric being the sum of static
spherically symmetric one and a small nondiagonal component
which describes rotation sufficiently slow to be taken into
account in the linear approximation. We study semiclassical
Einstein equations for this metric with vacuum expectation
value of stress-energy of physical fields as the source. In
suggestion that the static traversible wormhole solution
exists we reveal possible azimuthal angle dependence of
angular velocity of the rotation (angular velocity of the
local inertial frame) that solves semiclassical Einstein
equations. We find that in the macroscopic (in the Plank
scale) wormhole case a rotational solution exists but only
such that, first, angular velocity depends on radial
coordinate only and, second, the wormhole connects the two
asymptotically flat spacetimes rotating with angular
velocities different in asymptotic regions.
\end{abstract}
\newpage
{\bf 1.Introduction.} The possibility of existence of
static spherically-symmetrical \\ traversible wormhole as
topology-nontrivial solution to the Einstein equations has
been first studied by Morris and Thorne in 1988 \cite{MT}.
Since that time much activity has been developed in
studying the wormhole subject (see, e.g., review by Visser
\cite{Vis}). Rather interesting is the possibility of
existence of self-consistent wormhole solutions to
semiclassical Einstein equations. Checking this possibility
requires finding vacuum expectation value of the
stress-energy tensor as functional of geometry and solving
the Einstein equations with quantum backreaction, i.e. with
such the induced stress-energy as a source.

Recently some arguments in favour of this possibility has
been given. In Refs. \cite{Kh1, Kh2, Kh3} the gravity
induced vacuum stress-energy tensor in the wormhole
background has been found to violate energy conditions just
as it is required for this tensor itself be the source for
such the wormhole metric \cite{MT, MTY}. (We consider
physical vacuum of spin 1 and 1/2 massless fields in these
papers). In Ref. \cite{HPS} self-consistent
spherically-symmetrical wormhole solution has been found
numerically for the quantised scalar field vacuum playing
the role of a source for gravitation.

The problem of existence of self-consistent static wormhole
is a particular case of the more fundamental problem of
self-consistent solutions to semiclassical Einstein
equations with vacuum expectations of stress-energy tensor
of physical fields as a source. In Ref. \cite{Hor} it has
been found that such the problem linearised over metric
perturbations off Minkowski spacetime gives solutions
(apart from some unphysical ones) coinciding with those for
classical gravity wave problem.  If, however, topology is
not Minkowski one as in the wormhole case at hand, new
solutions can appear such as static wormhole itself. The
natural next step may be the search for stationary but
nonstatic topology nontrivial solution, namely rotating
vacuum wormhole. In the case of slow rotation one simply
adds small nondiagonal polar-angle-time component of metric
to the self-maintained static wormhole metric:
\begin{equation}
\label{ds-2}                                             
ds^2=\exp{(2\Phi)}dt^2-d\rho^2-r^2[d\theta^2
+\sin^2{\!\theta}
\,(d\phi^2+2hd\phi dt)]
\end{equation}
where $h(\rho ,\theta)$ has the sense of angular velocity
of the local inertial frame and will be called the angular
velocity of rotation in what follows. The static metric
$\Phi (\rho )$, $r(\rho )$ is assumed to exist as a
self-consistent solution of static semiclassical Einstein
equations, and solution for $h$ is to be found. One can
take $h$ arbitrarily small in order to limit oneself to the
theory linearised in $h$. In practice, in order that the
quadratic in $h$ terms in the Riemann tensor could be
disregarded, the space derivatives of $h$ should be
negligible as compared to the derivatives of the static
part of metric in the scale of typical wormhole size. Note
that in the linear approximation the only component of
Einstein equations being new compared to the static case is
the $t\phi$ one, as it follows from symmetry
considerations. The role of the source is played there by
the induced vacuum energy flow (angle-time stress-energy
component).

In the given note we show that the rotational solution
exists at least in the case when the coefficient at the
Weyl term in the induced stress-energy is large and if one
can neglect other terms. As noted in Refs. \cite{Kh2, Kh3},
this corresponds to the wormhole of macroscopic size (that
is, large in Plank units).  The rotational solution which
we argue should exist is such that $h$ depends on radial
distance $\rho$ only and has different asymptotic limits in
the two asymptotically flat spacetimes connected by
wormhole. This, in particular, means that these two
spacetimes cannot be "glued" together in asymptotic region,
so that this wormhole cannot be considered as that
connecting the two regions of {\it the same} asymptotically
flat spacetime.

As for the general case when we do not assume the Weyl term
to dominate, we find the two kinds of azimuthal angle
$\theta$ dependence of angular velocity $h$ for which the
radial and angular variables $\rho$, $\theta$ are separated
and the Einstein equation with quantum backreaction for $h$
reduces to that for the function of purely $\rho$. One
possibility is the above mentioned angle-independent angle
velocity; another one is proportional to $\cos{\theta}$
velocity.

\bigskip
{\bf 2.Classical rotation.} By classical we mean rotation
considered without taking into account corresponding
quantum backreaction, i.e. induced vacuum energy flow.
However, it is implied that the static wormhole problem is
already solved (the static metric is found) with taking
into account corresponding backreaction. We show (in the
linearised in $h$ theory used throughout the paper) that
only solution for $h$ not depending on $\theta$ is
physically acceptable such that it has different finite
limits at $\rho\rightarrow +\infty$ and at $\rho\rightarrow
-\infty$.

The $t\phi$ Einstein equation of interest can be
conveniently written using the tetrad components introduced
like the following basic 1-forms $\omega^a
=e^a_{\mu}dx^{\mu}$ \cite{Chan}:
\begin{equation}
\label{omega}                                            
\omega ^0=\exp{(\Phi)}dt,~~\omega ^1=(d\phi +hdt)r
\sin{\theta},~~
\omega ^2=d\rho,~~\omega ^3=rd\theta.
\end{equation}
Taking the expressions for the Riemann tensor presented in
Ref. \cite{Chan} we find for the equation of interest:
\begin{eqnarray}
\label{R-10}                                             
-R_{10}\left (=-{\exp{(\Phi)}\over r\sin{\theta}}R^t_{\phi}
\right )={1\over 2r^3}{\partial\over\partial\rho}
\exp{(-\Phi)}r^4
{\partial h\over\partial\rho}\sin{\theta}\nonumber\\
+{1\over 2r}\exp{(-\Phi)}
{1\over\sin^2{\theta}}{\partial\over\partial\theta}
\sin^3{\theta}{\partial h\over\partial\theta}=0.
\end{eqnarray}
Remind that we neglect the induced vacuum energy flow
$T^t_{\phi}$ in this section. Separating the variables,
$h=f(\rho)Y(\theta)$, we find hypergeometric function for
the angle dependence:
\begin{equation}
\label{F}                                                
Y\sim F(a,3-a;2;{1-\cos{\theta}\over 2}),
\end{equation}
$a=const$. This function diverges at $\theta =\pi$ unless
$a=-k$, $k=0,1,2,...$; in the latter case it reduces to the
Gegenbauer polynomial $C^{(3/2)}_k(\cos{\theta})$. Then the
radial part satisfies the equation
\begin{equation}
\label{class-rad}                                        
(r^4\exp{(-\Phi)}f^{\prime})^{\prime}=k(k+3)r^2
\exp{(-\Phi)}f.
\end{equation}
By asymptotical flatness $\Phi\rightarrow 0$,
$r\rightarrow\rho$ at $\rho\rightarrow\pm\infty$ and thus
$f\sim\rho ^{-k-3}$ at large distances. Another solution,
$f\sim\rho ^k$, should be discarded at $k\geq 1$ as
unphysical one. Therefore if we choose some large $L>0$, we
shall have solutions at $|\rho|>L$ parametrised by two
constants $C_{\pm}$ ($f\rightarrow C_{\pm}
\rho ^{-k-3}$ at $\rho\rightarrow\pm\infty$), whereas in the
intermediate region $|\rho|<L$ the Eq. (\ref{class-rad}) is
regular in the wormhole geometry and has a solution
parametrised by two constants $C_1$, $C_2$. The four
constants $C_1$, $C_2$, $C_+$, $C_-$ are subject to four
uniform equations which are matching conditions for $f$ and
it's derivative $f^{\prime}$ at $\rho =+L$ and at $\rho
=-L$. Requiring for this system to have nonzero solution,
i.e. zero determinant, we get some constraint on the
already known metric functions $r(\rho)$, $\Phi (\rho)$.
Therefore the set of possible solutions for $f(\rho)$ at
$k\geq 1$ has zero measure as compared to the set of
possible static solutions $r$, $\Phi$. At $k=0$ the
Eq. (\ref{class-rad}) can be easily integrated and leads to
physically admissible $h$ not depending on $\theta$ and
being monotonic function of $\rho$ which has finite but
different limits at $\rho\rightarrow\pm
\infty$: $h(-\infty)\neq h(+\infty)$. Because of the latter
circumstance the two asymptotically flat spacetimes
connected by the wormhole channel cannot be "glued"
together in asymptotic region (glueing at a shorter
distance would spoil spherical symmetry of the background
static solution described by two functions $r$, $\Phi$) so
we cannot derive from such the rotating wormhole the
wormhole connecting the two distinct regions of {\it the
same} asymptotically flat spacetime.

\bigskip
{\bf 3.Quantum backreaction and angle dependence of
rotation.} Now consider possible dependence of the angle
velocity $h$ on the azimuthal angle $\theta$ that could
solve the semiclassical Einstein equation (with
backreaction). Here we show that the only two versions of
the angle dependence for this equation to be solved by
separation of variables are the following ones: $h=f(\rho)$
or $h=f(\rho)\cos{\theta}$.

Evidently, the problem reduces to studying the angle
dependence of $T_{10}$ for a given angle dependence of $h$.
Given any physical field, we should solve equations of
motion for this field in curved spacetime and sum vacuum
contributions into stress-energy from all the eigenmodes.
The resulting expression can be regularised by, e.g.,
covariant geodesic point separation and renormalised by
subtracting the divergent parts known for physical fields
\cite{Christ}. Choosing such separation in the radial
direction we avoid discussing the renormalisation issue as
far as the angle dependence is concerned.

Let us consider general structure of the equations of
motion and stress-energy for arbitrary field in the metric
(\ref{ds-2}) and illustrate this by the case of massless
fields of spin 1 (electromagnetic) and 1/2 (neutrino). Most
natural to display effect of rotation in the stationary
axisymmetrical nonstatic metric is to use the complex
Newman-Penrose formalism in analogy with that applied to
Kerr metric \cite{Chan}. In particular, up to the linear
order in $h$, we can choose isotropic tetrad of real
$l^{\mu}$, $n^{\mu}$ and complex $m^{\mu}$, $m^{*\mu}$
(asterics means complex conjugation) of which $l^{\mu}$,
$n^{\mu}$ are tangential to some geodesics and $m^{\mu}$
(and thus $m^{*\mu}$) are orthogonal to $l^{\mu}$,
$n^{\mu}$. Besides that, normalisation can be chosen such
that $l_{\mu}n^{\mu}=1$, $m^{\mu}m^*_{\mu}=-1$. These are
the distinctive properties of the Newman-Penrose tetrad
which can be written with the help of the derivatives over
directions as
\begin{eqnarray}
\label{e-mu-d-mu}                                        
&&l^{\mu}\partial _{\mu}=\exp{(-2\Phi)}[\partial _t
+\exp{(\Phi)}\partial _{\rho}-h\partial
_{\phi}],\nonumber\\ &&n^{\mu}\partial _{\mu}={1\over
2}[\partial _t -\exp{(\Phi)}\partial _{\rho}-h\partial
_{\phi}],\\ &&m^{\mu}\partial
_{\mu}=(r\sqrt{2})^{-1}[\partial _{\theta}
+i(\sin{\theta})^{-1}\partial _{\phi}],\nonumber\\
&&m^{*\mu}\partial _{\mu}=(r\sqrt{2})^{-1}[\partial
_{\theta} -i(\sin{\theta})^{-1}\partial _{\phi}].\nonumber
\end{eqnarray}
Calculation gives the following values for the standard 12
complex spin-connection coefficients $\kappa$, $\nu$,
$\sigma$, $\lambda$, $\varepsilon$, $\varrho$, $\mu$,
$\tau$, $\pi$, $\alpha$, $\beta$, $\gamma$ \cite{Chan}:
\begin{eqnarray}
\label{connect}                                          
&&\kappa=\nu=0,~~~
\sigma=2\exp{(-2\Phi)}\lambda=-2\varepsilon={i\over 2}
\exp{(-2\Phi)}h_{\theta}\sin{\theta},\nonumber\\
&&\varrho=2\mu=-{r^{\prime}\over r}\exp{(-\Phi)},~~~
-\tau=\pi={i\over 2\sqrt{2}}\exp{(-\Phi)}h_{\rho}r
\sin{\theta},\\
&&-\alpha=\beta={1\over 2\sqrt{2}}{\cot{\theta}\over r}
-{i\over 4\sqrt{2}}\exp{(-\Phi)}h_{\rho}r\sin{\theta},~~~
\gamma={1\over 2}\Phi^{\prime}\exp{(\Phi)}-{i\over 8}
h_{\theta}\sin{\theta}\nonumber
\end{eqnarray}
where subscript on $h$ means corresponding derivative;
prime on $r$, $\Phi$ means derivative over $\rho$. The main
feature of the covariant equations of motion for an
arbitrary physical field is therefore occurence of $h$ in
the form $h_{\rho}
\sin{\theta}$, $h_{\theta}\sin{\theta}$ and $h\partial
_{\phi}$ there. In the diagrammatic language,
$h$-field-field vertex is combination of these expressions.
Besides that, in the Newman-Penrose formalism operators
acting on the angle variables appear in the form of spin
raising and lowering operators
\begin{equation}
\label{L-s}                                              
{\cal L}_s=\partial _{\theta}-{i\partial
_{\phi}\over
\sin{\theta}}+s\cot{\theta},~~~
{\cal L}^+_s=\partial _{\theta}+{i\partial _{\phi}\over
\sin{\theta}}+s\cot{\theta},
\end{equation}
$0\leq s\leq s_0$, $s_0$ being spin of the field. These
operators simplify in the basis of spin spherical harmonics
proportional to the elements of rotation matrix
$D^l_{sm}(\phi,\theta,0)$ \cite{Gold}: ${\cal L}_s$ and
${\cal L}^+_s$ transform $D^l_{sm}$ to $D^l_{s+1,m}$ and
$D^l_{s-1,m}$, respectively. In this basis the expression
for $T_{10}$ turns out to be a value of spin weight $s=\pm
1$ (and $m=0$), that is, combination of $D^l_{10}$ (or
$D^l_{-1,0}$) for different $l$. Quantum contribution to
$T_{10}$ can be viewed as some loop diagram with external
$T_{10}$- and $h$-legs. Since $h$ can be expanded in the
Legendre polynomials $P_k\sim D^k_{00}$, we can take
$D^k_{00}$ as probe function for the angle dependence of
$h$. Then the expressions $h_{\rho}\sin{\theta}$ and
$h_{\theta}\sin{\theta}$ appearing in the vertex on
$h$-line are combinations of the values $D^{k+1}_{10}$ and
$D^{k-1}_{10}$.  This corresponds to the angular momenta
$k+1$ and $k-1$ flowing through the $h$-line. It is less
evident but shown at the end of this section that the
vertex $h\partial _{\phi}$ corresponds to the combination
of angular momenta $k+1$, $k-1$, $k-3$, ... . By
conservation of angular momentum the $T_{10}$ also should
be combination of $D^{k+1-2n}_{10}$, $n=0,1,2$, ... . In
particular, at $k=0,1$ the only term ($D^1_{10}$ or
$D^2_{10}$) remains and $T_{10}$ factorises into the
functions of $\rho$ and of $\theta$. Moreover, since
$D^{k+1}_{10}\sim C^{3/2}_k$ just for $k=0,1$, the same
$\theta$-dependence also factors out in the LHS of Einstein
equation. Therefore we conclude: $h=f(\rho)$ or $h
=f(\rho)\cos{\theta}$ solves for the $\theta$-dependence of
semiclassical Einstein equation.

Finally, let us illustrate the above said by the examples
of electromagnetic and neutrino fields; for more detail on
the Newman-Penrose description of these fields see
Ref. \cite{Chan}.  Electromagnetic field is described by
three complex functions $f_0$, $f_1$, $f_2$; for our choice
of complex tetrad (\ref{e-mu-d-mu}) these are related to
the electromagnetic field strength tensor $F_{\mu\nu}$ as
follows:
\begin{eqnarray}
\label{strength}                                         
&&2f_0=F_{t\theta}+hF_{\theta\phi}+\exp{(\Phi)}
F_{\rho\theta}
+[F_{t\phi}+\exp{(\Phi)}F_{\rho\phi}]{i\over\sin{\theta}},
\nonumber\\
&&2f_1=-r^2\exp{(-\Phi)}(F_{t\rho}+hF_{\rho\phi})
+F_{\theta\phi}
{i\over\sin{\theta}},\\
&&2f_2=-(F_{t\theta}+hF_{\theta\phi})+\exp{(\Phi)}
F_{\rho\theta}
+[F_{t\phi}-\exp{(\Phi)}F_{\rho\phi}]{i\over\sin{\theta}}.
\nonumber
\end{eqnarray}
The eight real Maxwell equations can be recast into the
following four complex ones:
\begin{eqnarray}
\label{Maxwell}                                         
&&\partial _-f_1-{\cal L}_1f_0=h\partial_{\phi}f_1,
\nonumber\\
&&\partial _+f_1+{\cal L}^+_1f_0=-h\partial_{\phi}f_1,\\
&&\partial _+f_0+{\exp{(2\Phi)}\over r^2}{\cal L}^+_0f_1
=i{f_0-f_2\over 2}h_{\theta}\sin{\theta}
-ih_{\rho}\exp{(\Phi)}
\sin{\theta}f_1-h\partial _{\phi}f_0,\nonumber\\
&&\partial _-f_2-{\exp{(2\Phi)}\over r^2}{\cal L}_0f_1
=-i{f_0-f_2\over 2}h_{\theta}\sin{\theta}
+ih_{\rho}\exp{(\Phi)}
\sin{\theta}f_1+h\partial _{\phi}f_2,\nonumber
\end{eqnarray}
where $\partial _{\pm}\equiv\exp{(\Phi)}\partial _{\rho}\mp
\partial _t$, $i\partial _t=\omega$ being energy. Each mode
should be normalised so that it's full energy
\begin{eqnarray}
\label{norma-em}                                        
\int{T^t_tr^2\exp{(\Phi)}d\rho\sin{\theta}d\theta d\phi}
=\int{\sin{\theta}d\theta d\phi\exp{(-\Phi)}d\rho\Biggl\{
f^*_0f_0+f^*_2f_2\Biggl.}\nonumber\\
\left. +2{\exp{(2\Phi)}\over r^2}f^*_1f_1
-2h\sin{\theta}{\cal I}m[(f_0-f_2)^*f_1]\right\}
\end{eqnarray}
be equal to the vacuum value $\omega /2$ and then
substituted into the expression for the stress-energy
component studied,
\begin{equation}
\label{T-10-em}                                         
T_{10}\left (={\exp{(\Phi)}\over r\sin{\theta}}T^t_{\phi}
\right )=-2{\exp{(-\Phi)}\over r^3}
{\cal I}m[(f_0-f_2)^*f_1].
\end{equation}
Analogously, massless fermion field is described by two
complex values $g_1$, $g_2$ obeying the field equations
\begin{eqnarray}
\label{neutr-eqs}                                       
\partial _-g_1+{\exp{(\Phi)}\over r}{\cal L}_{1/2}g_2
={i\over 4}\sin{\theta}(g_1h_{\theta}-rh_{\rho}g_2)
+h\partial _{\phi}g_1,\nonumber\\
{\exp{(\Phi)}\over r}{\cal L}^+_{1/2}g_1-\partial _+g_2
={i\over 4}\sin{\theta}(-rh_{\rho}g_1-g_2h_{\theta})
+h\partial _{\phi}g_2.
\end{eqnarray}
Expression for the energy of each mode takes the form (on
the field equations)
\begin{equation}
\label{norma-neutr}                                     
\int{T^t_tr^2\exp{(\Phi)}d\rho\sin{\theta}d\theta d\phi}
=\int{\sin{\theta}d\theta d\phi\exp{(-\Phi)}d\rho\cdot
4\omega (g^*_1g_1+g^*_2g_2)}
\end{equation}
while the stress-energy component of interest is
\begin{eqnarray}
\label{T-10-neutr}                                      
T_{10}={\exp{(-\Phi)}\over r^3}\left\{g^*_1{\cal L}^+_{1/2}
g_1+({\cal L}^+_{1/2}g_1)^*g_1-g^*_2{\cal L}_{1/2}g_2
-({\cal L}_{1/2}g_2)^*g_2-\partial _{\theta}
(g^*_1g_1-g^*_2g_2)\nonumber\right.\\
\left.+(r\Phi ^{\prime}-r^{\prime})(g^*_2g_1+g^*_1g_2)
+2r\exp{(-\Phi)}[g^*_2(\partial _t-h\partial _{\phi})g_1
-g^*_1(\partial _t-h\partial _{\phi})g_2]\right\}.
\end{eqnarray}

Solutions to the above equations of motion can be
constructed iteratively. In zero order one takes
\begin{equation}
\label{f-zero}                                          
\left (\matrix{f_0\cr f_1\cr f_2\cr }\right )^{(0)}
=\left (\matrix{\partial _-RD^l_{-1,m}\cr
-\sqrt{l(l+1)}RD^l_{0m}\cr \partial _+RD^l_{+1,m}\cr}\right
)
\end{equation}
which are the well-known TE-modes for the electromagnetic
field (TM-modes follow by multiplying this by
$i=\sqrt{-1}$) and
\begin{equation}
\label{g-zero}                                          
\left (\matrix{g_1\cr g_2\cr }\right )^{(0)}=\left (
\matrix{Z_1D^l_{+1/2,m}\cr Z_2D^l_{-1/2,m}\cr }\right )
\end{equation}
for the neutrino field. The $R$ and $Z_1$, $Z_2$ are some
radial functions. Substituting Eqs. (\ref{f-zero}) and
(\ref{g-zero}) into the RHS of the equations of motion we
find for the first $O(h)$ correction an expression of the
type
\begin{eqnarray}
\label{f-first}                                         
\left (\matrix{f_0\cr f_1\cr f_2\cr }\right )^{(1)}
=\sum_j{i(-1)^m\left [\left (\matrix{\dots D^j_{-1,m}\cr
\dots D^j_{0m}\cr \dots D^j_{+1,m}\cr }\right )\left (
\matrix{l~k-1~~j\cr m~~0~-m\cr }\right )
+\left (\matrix{\dots D^j_{-1,m}\cr
\dots D^j_{0m}\cr \dots D^j_{+1,m}\cr }\right )\times
\right. } \nonumber\\
\left. \left (\matrix{l~k+1~~j\cr m~~0~-m\cr }\right )
+\left (\matrix{\dots D^j_{-1,m}\cr
\dots D^j_{0m}\cr \dots D^j_{+1,m}\cr }\right )m\left (
\matrix{l~~k~~~~~j\cr m~0~-m\cr }\right )\right ]
\end{eqnarray}
and quite analogous one for the fermion field. The dots
mean (real) factors which do not depend on $\theta$, $\phi$
and $m$.  The linear order in $h$ of interest comes from
interplay in the bilinear $T_{10}$ between zero order
(\ref{f-zero}) and first correction (\ref{f-first}). By
properties of 3j-symbols and rotation matrix elements
$D^l_{sm}$ summation over $m$ just yields combination of
the harmonics $D^{k+p}_{10}$, $p=1,-1,-3,
\dots $. Important is that the last term in (\ref{f-first})
which stems from $h\partial _{\phi}$ operator in the
equations of motion is representable as combination of
3j-symbols as
\begin{equation}
\label{m-3j}                                            
m\left (\matrix{l~~k~~~~~j\cr m~0~-m\cr }\right )
=\sum_{n\geq 0}{A_n\left (\matrix{l~~k+1-2n~~j\cr
 m~~~~~0~~~~~-m\cr }
\right )}
\end{equation}
where $A_n$ does not depend on $m$. In $T_{10}$ this and
other terms enter multiplied by $D^l_{0m}(D^j_{\pm 1,m})^*$
and summed over $m$ thus giving just combination of
$D^{k+1-2n} _{10}$, $n\geq 0$.

\bigskip
{\bf 4.Macroscopic wormhole and radial dependence of
rotation.} Usually, if one does not assume existence of
fundamental scales in the theory other than the Plank scale
one expects the typical wormhole size be of the Plank scale
too. However, a new scale can exist connected with
coefficient of the Weyl term in the effective action. This
coefficient is subject to renormalisation in both infrared
(if massless fields are present in the theory) and
ultraviolet regions. Possible large value of this
coefficient can enable existence of the wormhole of
macroscopic size.

Here we argue that if the Weyl term coefficient is large
and one can disregard other terms in the effective action
then the conclusion concerning the existence of rotating
wormholes resembles that for the case of classical rotation
in Sect. 2.

The effective gravity Lagrangian density with taking into
account the Weyl term can be written as proportional to $R
+(2\mu^2)^{-1}C_{\mu\nu\lambda\rho}C^{\mu\nu\lambda\rho}$
\cite{DeWitt}. Up to the full derivative, the Weyl term
$C_{\mu\nu\lambda\rho}C^{\mu\nu\lambda\rho}$ is equivalent
to $2(R_{\mu\nu}R^{\mu\nu}-{1\over 3}R^2)$. We calculate
the latter up to the second order in $h$ (required to get
the first order in the equations of motion) using Riemann
tensor given in Ref. \cite{Chan} in the tetrad components.
Varying in $h$ gives the desired $t\phi$-component of the
Einstein equations.  Consider both versions of
$\theta$-dependence of $h$ found in Sect. 3, $h\sim 1$
and $h\sim\cos{\theta}$, and introduce new variable $z$ via
$dz=\exp{(-\Phi)}d\rho$ and the function
$\tilde{r}=r\exp{(-\Phi)}$. For $h$ not depending on
$\theta$, $h=f(\rho)=f(\rho (z))$, the result reads
\begin{equation}
\label{10-weyl-1}                                       
(\tilde{r}^4f_z\exp{(2\Phi)})_z={1\over\mu ^2}
\left \{\tilde{r}^4\left [{1\over\tilde{r}^4}(
\tilde{r}^4f_z)_z
\right ]_z+\left [{10\over 3}(\tilde{r}^2_z-1)+{2\over 3}
\tilde{r}_{zz}\tilde{r}\right ]\tilde{r}^2f_z\right \}_z
\end{equation}
(subscript $z$ means differentiation over $z$). For
$h=f(\rho)\cos{\theta}$ we find
\begin{eqnarray}
\label{10-weyl-cos}                                     
(\tilde{r}^4f_z\exp{(2\Phi)})_z-4\exp{(2\Phi)}\tilde{r}^2f
={1\over\mu ^2}\left \{\left [\tilde{r}^4\left (
{1\over\tilde{r}^4}(\tilde{r}^4f_z)_z\right )_z
\right.\right.\nonumber\\
\left.\left.+\left ({10\over 3}\tilde{r}^2_z+{2\over 3}
\tilde{r}_{zz}\tilde{r}-{34\over 3}\right )
\tilde{r}^2f_z\right ]_z
+{8\over 3}(-\tilde{r}_{zz}\tilde{r}
+\tilde{r}^2_z+8)f\right \}.
\end{eqnarray}
The infrared contribution to the coefficient $\mu ^{-2}$
goes from the massless fields. In the considered case of
$\mu ^{-2}$ large in the Plank scale the typical wormhole
size $r^2_0$ is defined just by $\mu ^{-2}$ as $r^2_0=(3\mu
^2)^{-1}$ \cite{Kh2, Kh3}. For example, in the vacuum of
$N_1$ spin 1 and $N_{1/2}$ spin 1/2 massless fields we have
\begin{equation}
\label{size}                                            
(3\mu ^2)^{-1}=r^2_0={G\over 120\pi}(4N_1+N_{1/2})
\ln{\left ({120\pi\over G}{\Lambda ^2\over 4N_1+N_{1/2}}
\right )},
\end{equation}
$\Lambda$ being infrared cut off.

Consider first Eq. (\ref{10-weyl-1}) which upon integrating
both parts and denoting $f_z\equiv g$ reduces to the second
order one:
\begin{equation}
\label{g-eq}                                            
\mu ^2\exp{(2\Phi)}\tilde{r}^4g+C=\tilde{r}^4\left [
{1\over\tilde{r}^4}(\tilde{r}^4g)_z
\right ]_z+\left [{10\over 3}(\tilde{r}^2_z-1)+{2\over 3}
\tilde{r}_{zz}\tilde{r}\right ]\tilde{r}^2g
\end{equation}
where $C=const$. Asymptotical $\rho\rightarrow\pm\infty$
form of this equation reads
\begin{equation}
\label{g-asymp}                                         
{d\over d\rho}{1\over\rho ^4}{d\over d\rho}\rho ^4g-\mu ^2g
={C\over\rho ^4}.
\end{equation}
The general solution is the sum of a particular one which
behaves as $g\sim\rho ^{-4}+O(\rho ^{-6})$ at
$\rho\rightarrow\pm\infty$ and arbitrary combination of the
two independent solutions to the uniform equation. Of the
latter two one exponentially grows at $\rho\rightarrow
+\infty$ or at $\rho\rightarrow -\infty$ and should be
omitted as unphysical solution while another one
proportional to $\exp{(-\mu\rho)}$ (at $\rho\rightarrow
+\infty$) or $\exp{(\mu\rho)}$ (at $\rho\rightarrow
-\infty$) should be kept. Therefore, if we choose, as in
Sect. 2, some large $L>0$ we shall have physically
acceptable solutions at $|\rho|>L$ parametrised by three
constants, one of which is $C$. Meanwhile, in the
intermediate region $|\rho|<L$ the equation (\ref{g-eq}) is
regular in the wormhole geometry and has solution
parametrised by maximal set of three constants, one of
which is $C$. The overall set of five constants is subject
to four uniform equations which are matching conditions for
$g$ and for it's derivative at $\rho =+L$ and at $\rho
=-L$. This defines all five constants up to an overall
factor. Note that imposing additional condition
$\int^{+\infty}_{-\infty}{gdz}=0$ (that is, $h(-\infty )
=h(+\infty )$) is, generally speaking, contradictory since
it would be condition not on a freely chosen constant, but
on the already defined static metric $\Phi (\rho)$,
$r(\rho)$.

Next consider Eq. (\ref{10-weyl-cos}) which has asymptotic
form (at $\Phi =0$, $r=\rho$)
\begin{equation}
\label{f-asymp}                                         
\rho ^4{d\over d\rho}{1\over\rho ^4}{d\over d\rho}\rho ^4f
={1\over\mu ^2}\rho ^4\left ({d\over d\rho}{1\over\rho ^4}
{d\over d\rho}\rho ^4\right )^2f.
\end{equation}
Assuming this form at $|\rho|>L$ we find
\begin{equation}
\label{f-reduce}                                        
{d\over d\rho}{1\over\rho ^4}{d\over d\rho}\rho ^4f-\mu ^2f
=C^{\pm}_{-1}\rho +{C^{\pm}_{+4}\over\rho ^4}
\end{equation}
where $C^+_{-1}$, $C^+_{+4}$ ($C^-_{-1}$, $C^-_{+4}$) are
some constants which parametrise the solution at $\rho >L$
(at $\rho <-L$). To get physical solution we put
$C^{\pm}_{-1}=0$. Of the two solutions of uniform equation
we discard exponentially growing one in each region $\rho
>L$ or $\rho <-L$ and retain exponentially falling off
another solution. Thus, in each region $\rho >L$ or $\rho
<-L$ the solution is specified by two constants. At the
same time, the regular fourth order differential equation
has solution parametrised by four constants at $|\rho |<L$.
The overall number of constants is eight. These should
ensure validity of eight matching conditions for $f$,
$f^{\prime}$, $f^{\prime\prime}$ and
$f^{\prime\prime\prime}$ at $\rho =\pm L$. The determinant
of this uniform system should be zero. This imposes a
constraint on the already known static metric $\Phi$, $r$.
Therefore the subset of rotating wormhole solutions with
$h=f(\rho )\cos{\theta}$ should have zero measure w.r.t.
the set of spherically symmetrical static wormhole
solutions $\Phi (\rho )$, $r(\rho )$.

\bigskip
{\bf 5.Conclusion.} We have shown that if the coefficient
at the Weyl term is large (infrared cut off is large) and
one can discard other terms in the effective action then
the rotation existing for any static wormhole background
$\Phi (\rho )$, $r(\rho )$ is that which proceeds with the
angular velocity $h$ not depending on the azimuthal angle
$\theta$ and having the different finite limits $h(+\infty
)$ and $h(-\infty )$ in the asymptotic region
$\rho\rightarrow\pm\infty$. We see that despite that the
structure of the equations is drastically changed because
of enhancing the maximal order of derivatives upon taking
into account vacuum polarisation, the result practically
does not differ from that for the classical case of
Sect. 2.

Note that the sign of infrared divergent coefficient $\mu
^{-2}$ at the Weyl tensor squared in the effective action
is crucial for existence (or, rather, nonexistence) of more
rotational solutions in our case. Were $\mu ^{-2}$
substituted by negative value, the equations above would
have oscillating instead of monotonic exponential
solutions, and we would not have to omit some of them as
unphysical ones. Then the general solution of interest
would be parametrised by more constants, and the set of
such solutions would be larger.

Also we can say that if we denote $\mu ^2\equiv x$ and
extend the equations to arbitrary real $x$, the problem
will be singular at $x=0$.

In the case if the infrared logarithm is not large we do
not have macroscopic vacuum wormhole, and the following
interesting question arises: whether microscopic wormhole
can rotate so that it would have the macroscopic "tail" of
rotation (when $h$ falls off in power law in asymptotic
region). Answering this question implies rather complicated
problem of calculation and analysis of the terms in
stress-energy other than the Weyl term.

\bigskip
This work was supported in part by the President Council
for Grants through grant No. 96-15-96317.

\end{document}